\documentclass[superscriptaddress,twocolumn,aps,prl,showpacs]{revtex4-1}
\usepackage{epsfig}
\usepackage{graphicx}
\usepackage{dcolumn}
\usepackage{bm}
\usepackage{amssymb}
\usepackage{subfigure}
\usepackage{float}
\usepackage{xcolor}

\begin{document}



\title{Collective learning from individual experiences and information transfer during group foraging}

\author{Andrea Falc\'on-Cort\'es}
\email{andreafalcon@estudiantes.fisica.unam.mx}
\affiliation{Instituto de F\'\i sica, Universidad Nacional Aut\'onoma de 
M\'exico, Ciudad de M\'exico 04510, M\'exico}

\author{Denis Boyer}
\email{boyer@fisica.unam.mx}
\affiliation{Instituto de F\'\i sica, Universidad Nacional Aut\'onoma de 
M\'exico, Ciudad de M\'exico 04510, M\'exico}

\author{Gabriel Ramos-Fern\'andez}
\affiliation{ Instituto de Investigaciones en Matem\'aticas Aplicadas y Sistemas, Universidad Nacional Aut\'onoma de M\'exico, Ciudad de M\'exico 04510, M\'exico}
\affiliation{Unidad Profesional Interdisciplinaria en Ingenier\'ia y Tecnolog\'ias Avanzadas, 
Instituto Polit\'enico Nacional, M\'exico}

\date{\today}

\begin{abstract}
Living in groups brings benefits to many animals, such as a protection against predators and an improved capacity for sensing and making decisions while searching for resources in uncertain environments. A body of studies has shown how collective behaviors within animal groups on the move can be useful for pooling information about the current state of the environment. The effects of interactions on collective motion have been mostly studied in models of agents with no memory. Thus, whether coordinated behaviors can emerge from individuals with memory and different foraging experiences is still poorly understood. By means of an agent based model, we quantify how individual memory and information fluxes can contribute to improving the foraging success of a group in complex environments. In this context, we define collective learning as a coordinated change of behavior within a group resulting from individual experiences and information transfer. We show that an initially scattered population of foragers visiting dispersed resources can gradually achieve cohesion and become selectively localized in space around the most salient resource sites. Coordination is lost when memory or information transfer among individuals is suppressed. 
The present modelling framework provides predictions for empirical studies of collective learning and could also find applications in swarm robotics and motivate new search algorithms based on reinforcement.

\end{abstract}

\maketitle

\section{Introduction}

Learning processes are of primary importance for many living organisms to adapt to their environments \cite{trimmer12, mcnamara09}. During foraging, animals can take movement decisions based on past successful experiences, resulting in changes of behaviors over time and an improved exploitation of resources \cite{bouton07, miller07}. The use of spatial memory is well documented in many species \cite{fagan13,morales14}, and ecological knowledge can be the result of continuous learning throughout the life of an individual. Large herbivores that were introduced into a novel environment took several years to adopt a phase of home-range movements \cite{moralespnas}. Adult seabirds like gannets have a better knowledge of profitable zones and forage more efficiently than young individuals \cite{grecian18}. Markovian random walk models, that assume that foragers lack memory, have proven useful to study how foraging success depends on particular movement patterns and the distribution of resources \cite{vis99, bart09}. However, such models do not consider the fact that animals perform some actions repeatedly and they cannot address the role played by experience on movement decisions \cite{sasaki13}. Recent mechanistic models have further incorporated memory-based movements, showing how spatial learning can emerge with time: learning is noticeable, for instance, by frequent revisits to certain locations rich in resources \cite{bonnell13,falcon17} or through the emergence of home ranges and preferred travel routes \cite{vanmoorter,boyer10}. Similar theoretical approaches have been useful to understand insect navigation \cite{cruse11, liho12}.\\

On the other hand, many animal species live in groups \cite{clark86,krause02} and one may ask how individual spatial learning may be affected, or enhanced, by the presence of others. { Sociality} brings many known benefits, such as a decreased risk of predation \cite{krause02,ioann11,ioann12,sumpter10} or improved capacities for sensing and making decisions \cite{berdahl13,mallon01,conradt05} while searching for food in uncertain environments \cite{girald00,pitcher82,haney92,girald99}. When individuals forage in groups, foraging success can be enhanced by information transfer between group members \cite{torney11, martinezgarcia13,dall05,lihoreau17}. { Social insects like honey bees \cite{seeley91} and ants \cite{bekers90} are classical examples of collective foraging where information is transferred between individuals by external means (chemical trails or nectar sharing). Similar principles can apply to vertebrates, where information is not necessarily stored in the environment but within the individual, such that individuals can use their internal memory to visit places located beyond their perceptual range.}  
Species such as macaques \cite{hauser93}, cliff swallows \cite{brown91}, gazelles \cite{frey08}, elephants \cite{mccomb03} and hyenas \cite{mathevan10} communicate over large distances (sometimes of several kilometers) by vocalizations. The information which is transferred in this way can have a positive impact on the foraging success in social groups \cite{frey08}. In killer whales, old females tend to lead the group during collective movement when prey abundance is low, thus transferring valuable ecological knowledge \cite{brent15}. In spider monkeys, central individuals in the social network tend to be followed more when they are knowledgeable about available feeding sources than other, non-central individuals \cite{palacios_inrev}.\\

Animal groups on the move, where individuals interact locally, are prone to exhibit collective behaviors \cite{krause02,couzin02,krause10,suro05} { through} which the group can accurately pool information about the current state of the environment \cite{berdahl13,frisch67,hoare04,couzin05}. Such \lq\lq collective intelligence" 
can allow a rapid response to external signals. The effects of interactions on collective motion \cite{krause10, sasaki17} and their impact on foraging success \cite{vicsek14} have been well studied in systems of agents without memory. { When} the {dynamical} rules are local in time, { they} only depend on the current state of the system and not on its previous history. 
In contrast, the mechanisms by which coordinated behaviors emerge from interacting individuals with memory are much less understood. { Modelling approaches have shown that a physiological memory can guide individual foraging behaviour and influence social interactions, ultimately affecting group dynamics \cite{hemelrijk,senior}.} { However, these former models focus on nutritional states rather than movement and space use.}
We wish to quantify here, by means of a { spatially explicit} agent based model, how both memory use and interactions with peers can contribute to improving the foraging success of a group as a whole.\\ 

We define collective learning as a coordinated change of behavior within a group, resulting from individual experiences and information transfer between group members {\cite{kao14}}. During collective learning, individuals with different experiences may acquire valuable information through interactions with others, possibly resulting in an increased foraging success compared to what isolated individuals would typically achieve. The structure of social networks is likely to be relevant in such processes, since certain individuals are more important than others for transmitting information on food locations \cite{brent15}.\\

We develop a simple agent based model to show that collective learning can { contribute significantly} to successful group foraging in complex environments, { namely, those} composed of { many} resource {sites of unequal values}, representing food or water holes. { When the agents can use memory to visit places located beyond their perceptual range}, we seek to determine whether { collective behaviours such as} spatial aggregation can emerge from { a} dispersed population with no initial information about resources, and whether such aggregation takes place around the most profitable { places}. It is also of our interest to study the effects of the foragers' interaction network on collective learning.\\

\section{Model}
The collective model presented here is an extension of another one for a single forager with reinforcement learning, exposed under slightly different forms in \cite{falcon17,gautestad05,gautestad06,boyersolis,boyerpineda}. To summarize, the motion of the individuals is assumed to be driven by the combination of two basic movement modes: standard random walk displacements and preferential returns to places visited in the past. In the latter mode, a forager (assuming first that it does not interact with others) chooses a particular site for relocation with a probability proportional to the accumulated amount of time it previously spent at that site. Therefore, the sites that are often occupied have a higher probability of being revisited in the future: they are linearly reinforced \cite{falcon17}. Since { these} preferential dynamics depend on history, it implies that the forager has the ability to remember which sites were previously visited and for how long. To incorporate interactions between several foragers, we assume here that memory is shared through constant communication. 
Hence, in the memory mode, a forager can relocate to a previously visited site chosen either from its own experience or from the experience of another group member. A second important aspect is the modeling of the resource landscape: each resource site is characterized by a weight or profitability $\gamma_i$. A forager located at a resource site $i$ stays on it at the next time step with probability $\gamma_i$ or moves with probability $1-\gamma_i$. These rules are detailed as follows.\\ 

{\bf Environment.} We consider {discrete lattices, of $L$ sites in one dimension ($1d$) and of $L\times L$ sites in two dimensions ($2d$),}  with unit spacing and reflective boundaries. The site coordinate $n$ is an integer ($1d$ case) or a pair of integers in $2d$. On the lattice, resources (or targets) are distributed randomly with density $\delta<1$, hence, on average, there are $M=\delta L$ targets. To each target $i$ is assigned a fixed weight $\gamma_i$, which is a random number uniformly distributed in the interval $(0,\gamma_{max})$, where $\gamma_{max}<1$ is a given parameter. Let $\mathbb{T}=\{(T_i,\gamma_i)\}_{i=1}^{M}$ denote the set that contains the positions $T_i$ and weights $\gamma_{i}$ of the targets.\\

{\bf Foragers.} We consider $N$ walkers with random initial positions on the lattice { and that are} connected by a { complete} communication network, namely, every walker is able to communicate with any other group member (see below for cases involving other interaction networks). The time variable $t$ is discrete. During a time step $t\rightarrow t+1$, each walker $l =1,...,N$ updates its position $\mathbf{X}^{(l)}_{t}$ as follows.

\begin{enumerate}
 \item [(\emph{i})] Self-memory mode: If not on a target, with probability $q(1-\rho)$ the walker resets to a site that it visited in the past, that is: $\mathbf{X}^{(l)}_{t+1} = \mathbf{X}^{(l)}_{t'}$ where $t'$ is a random integer uniformly chosen in the interval $[0,t]$. This implies that the probability of choosing a particular site (a target or not) is proportional to the number of time-steps previously spent on that site. This is the linear preferential revisit scheme mentioned above. 
\item [(\emph{ii})] Information transfer mode: If not on a target, with probability $q\rho$ the walker randomly chooses another walker ($m$) and relocates to a place already visited by that walker, according to the same preferential rule: $\mathbf{X}^{(l)}_{t+1}=\mathbf{X}^{(m)}_{t'}$ where $t'$ is a random integer in $[0,t]$. 
\item [(\emph{iii})] Random motion mode: If not on a target, with probability $1-q$ the walker moves to one of its nearest neighbor sites with equal probability each: $\mathbf{X}^{(l)}_{t+1} = \mathbf{X}^{(l)}_{t} +{\ell}_t$, where $\ell_t=\pm 1$ in one dimension.
 \item [(\emph{iv})] Feeding: If on a target (if there exists a $i$ such that $\mathbf{X}^{(l)}_{t}=T_\mathbf{i}$), the walker $l$ stays on that site with probability $\gamma_i$, and with the complementary probability $1-\gamma_i$ uses one of the movement rules $(\emph{i})-(\emph{iii})$ above with { their} respective probabilities.
\end{enumerate}

\begin{figure}
  \centerline{\includegraphics*[width=0.53\textwidth]{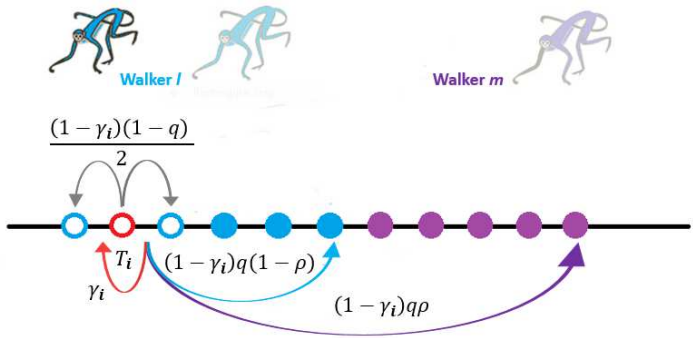}}
  \caption{Dynamics that consider feeding {[red arrow]}, random movement {[grey arrow]}, memory based-movement {[blue arrow]} and information transfer {[purple arrow]}, for a walker $l$ located on a resource site $T_{i}$. If not on a resource site, the same rules apply with $\gamma_{i}$ set to $0$.}
\label{figdynamic}
\end{figure}

Hence, in a time step, { the walker uses memory with probability $q$ and takes a random step with probability $1-q$}. If the memory mode is chosen, { with probability $1-\rho$, the walker relies on its own experience, and with probability $\rho$, on that} of another group member. The parameters $q$ and $\rho$ { are critical for learning}. Fig. \ref{figdynamic} { represents rules (\emph{i})-(\emph{iv}) schematically}.\\

{\bf Other interaction networks.} { We assume that walkers can obtain information from any other group member. This is realistic for those animal societies where there is a high fluidity in association patterns \cite{aureli_etal08}. Still, even these fluid association patterns may be represented by networks with different connectivity \cite{ramosfer09}.} In an arbitrary network, each walker (node $l$) has a fixed set of connections to other nodes. In this case, in rule $(\emph{ii})$ one of these nodes ($m$) is chosen at random for updating. As a representative example, we will further study the case of random Erd$\ddot{\mbox{o}}$s-Renyi (ER) networks, where a connection between any given pair of nodes exists with probability $s<1$ ($s=1$ corresponding to the complete graph) \cite{barabasi16,newman10}. We also explore the case of random scale-free networks, constructed with the method given in \cite{newman01}, where the number of nodes with $k$ connections is $N_k\propto k^{-\alpha}$. The power-law exponent $\alpha$ can take any value $>1$. { Scale-free networks are of interest because they contain highly connected nodes, a property not shared by ER networks. These nodes mimic central individuals, which play important roles in real animal social networks \cite{palacios_inrev, lusseau04,pinterwollman}. Scale-free animal social networks were discussed in \cite{kanngiesser}.}\\

{ The software specifications and source code with which we obtained most of the results can be found in \cite{githubjava}.}\\

\begin{figure}
\centerline{\includegraphics*[width=0.5\textwidth]{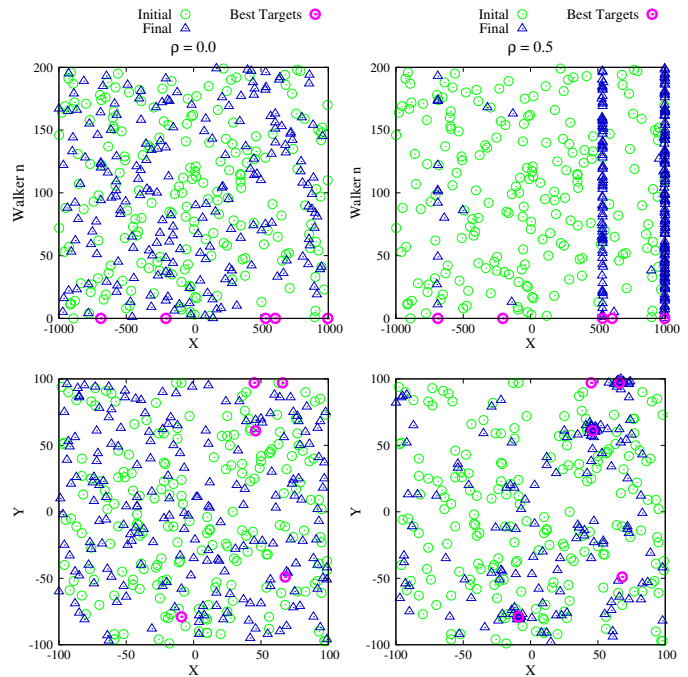}}
  \caption{Initial (green circles) and final (blue triangles) positions of 200 walkers after $t$ iterations of the dynamic rules (\emph{i})-(\emph{iv}). The five best targets (among $M$) are represented by pink circles. {In all cases, $q=0.1$ and $\gamma_{max}=0.9$. In the left column $\rho=0.0$ (no communication) and in the right, $\rho=0.5$.}. Upper panels:  $1d$ system of length $L=2000$ with $M=50$ targets, at $t=5\ 10^4$. Lower panels: $2d$ systems with $200\times200$ sites and $M=100$ targets, at $t=5\ 10^5$.}
\label{bestposition}
\end{figure}

Before proceeding to the results, we recall that the case $N=1$ and $M=1$ (a single forager and one target of weight $\gamma$), studied in \cite{falcon17}, exhibits an interesting localization phenomenon. If the rate of memory use $q$ is non-zero, the walker does not diffuse across the lattice but rather becomes localized for long times in a region centered around the target, independently of its initial position. Namely, because of the preferential return rule to the previously visited sites and because of the fact that each visit to the target lasts longer than those to any other site, a { subtle} reinforced learning process emerges where the walker steadily revisits the target site, which becomes the most occupied or \lq\lq preferred" site.

\begin{figure*}
    \centerline{\includegraphics*[width=0.85\textwidth]{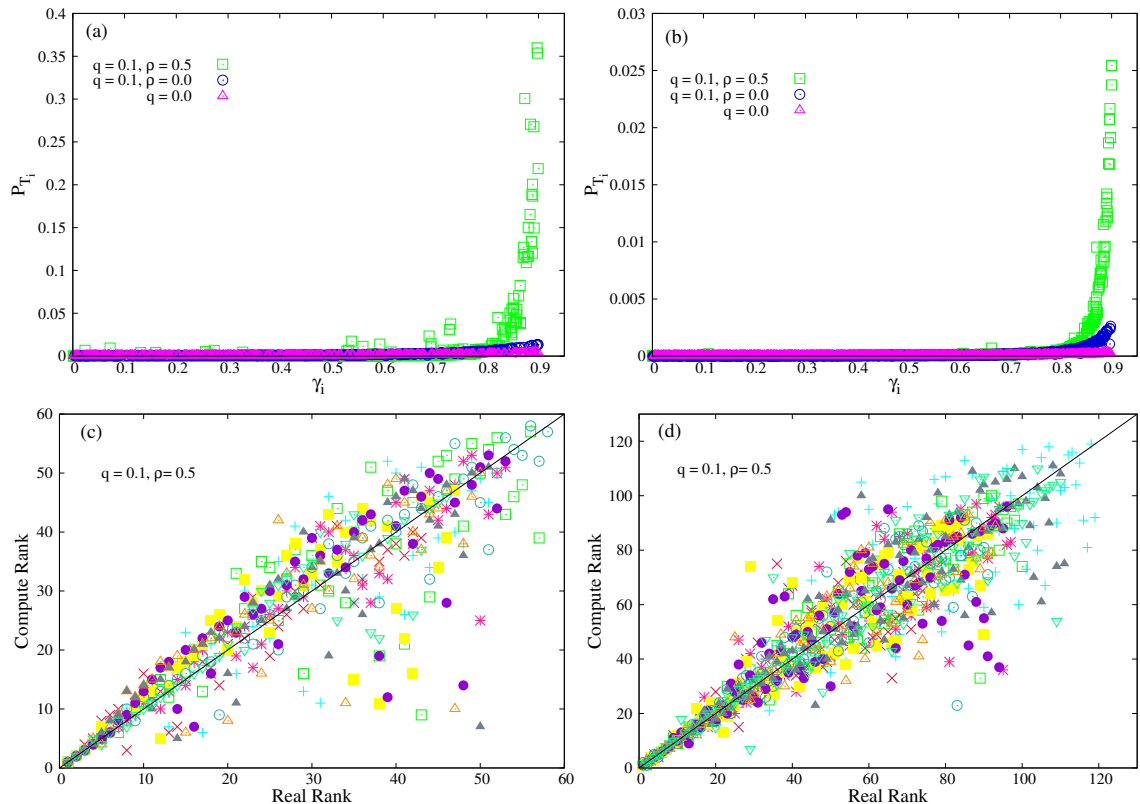}}
  \caption{{Simulations with $N=200$ walkers and $\gamma_{max}=0.9$}. (a){ $P_{T_i}$} in $1d$ vs. the value $\gamma_i$ of the corresponding target. Each point represents a target and $P_{T_i}$ is obtained from averaging over 1000 independent walkers dynamics in a same environment. The data from 10 different environments are aggregated. The other parameters are $t=5\ 10^4$, $L=2000$ and $\delta = 0.025$ ($M=50$). (b) Same quantity in a $2d$ space with 
$200\times 200$ sites and $\delta = 0.0025$ ($M=100$), at $t=5\ 10^5$. (c) Computed ranking of a target (based on its $P_{T_i}$) vs. its real ranking (based on its $\gamma_i$) for ten different $1d$ landscape configurations, with the same parameters as in (a). Each point represents a target in a particular environment and each symbol represents a different environment. (d) Same plot in $2d$, with the same parameters as in (b).}
\label{rankP}
\end{figure*}

\section{Results}

 Unless otherwise indicated, the results below are obtained for interactions on the complete graph. In Fig. \ref{bestposition} we show the walkers' initial and final positions (after $t$ time steps, with $t$ large) in { four} representative numerical simulations. The upper and lower panels correspond to one and two dimensional spaces, { and left and right panels are for $\rho=0$ and $\rho=0.5$}, respectively. In these figures we also display the positions of the targets with the five highest weights $\gamma_i$, { among 50 or 100 targets.} Notably, { for $\rho>0$}, many foragers tend to aggregate around these best targets, creating a much less uniform distribution than initially, { especially} in $1d$. { In contrast, when $\rho=0$, the forager distribution remains rather uniform}. 
This preference for occupying a few most profitable targets, among a large set of available sites and targets, can be referred to as \emph{selective localization}.\\

Resource selection by the walkers can be quantified through the final occupation probability of each target $T_i$ after a long simulation time, when a steady distribution is established: it is denoted as $P_{T_i}$ and is defined by the probability that a walker chosen at random occupies $T_i$. This quantity is represented against the target weight $\gamma_i$ in Figs. \ref{rankP} (a)-(b), where $\rho$ is set to $0.5$ ({ green squares}). One observes that $P_{T_i}$ is very small and practically independent of $\gamma_i$ in most of the interval $(0,\gamma_{max})$, whereas it sharply increases when $\gamma_i$ approaches its upper bound $\gamma_{max}$. In $1d$, there is a probability as high as $1/3$ to be on the target with the highest weight (among the $50$ targets and $2000$ total sites available). Our foragers are over-attracted by the most profitable sites and therefore very selective. In contrast, when the foragers do not interact ($\rho=0$, plots with blue circles), this selectivity practically disappears, despite the fact that the individuals still use their memory at the same rate $q$. In this case, each forager localizes around a target close to its initial position. { When the foragers are memory-less random walkers ($q=0$, pink triangles) the selectivity disappears altogether. In this case the foragers are not able to learn.} Therefore, the emergence of selective localization here is only possible in connected collectives { with memory}.\\


\begin{figure*}
    \centerline{\includegraphics*[width=1.0\textwidth]{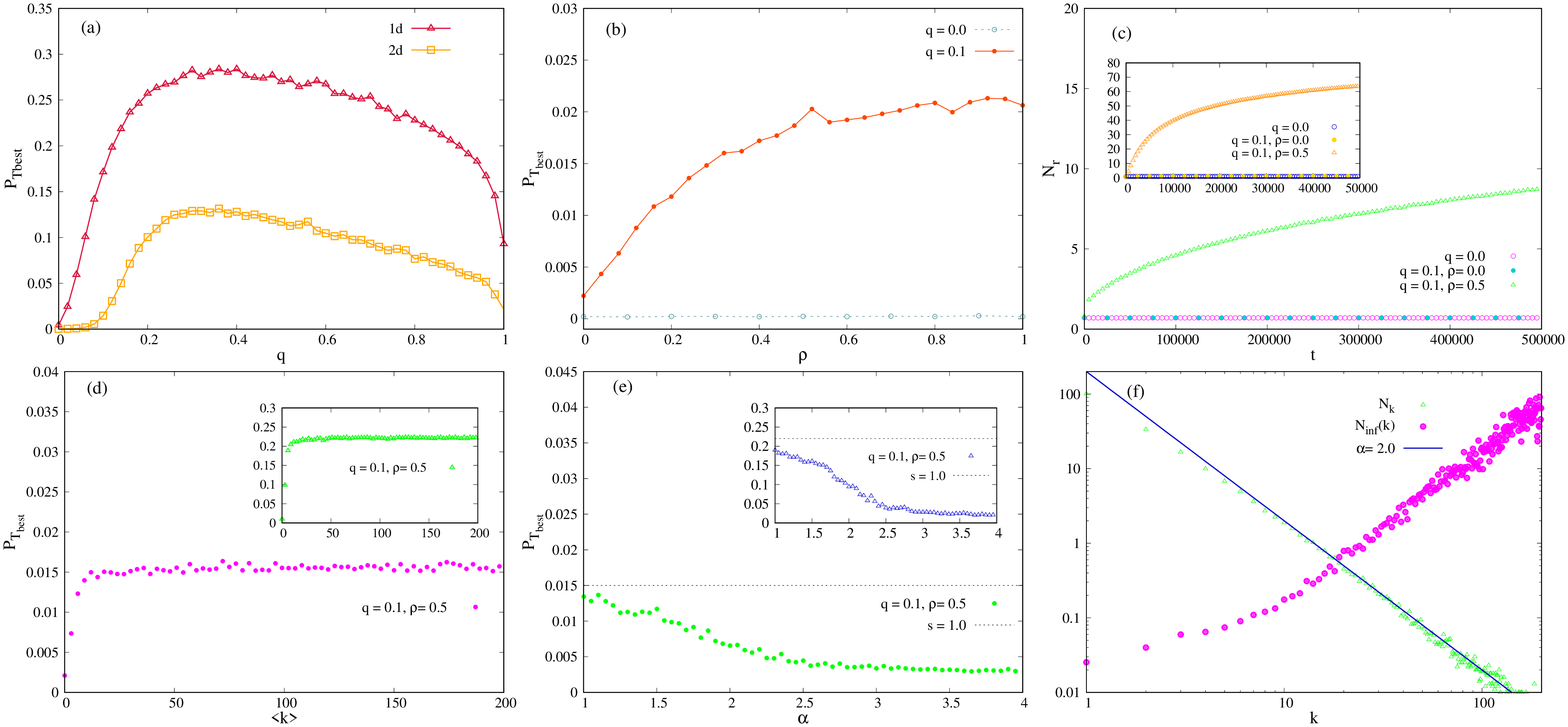}}
  \caption{{ Simulations with $N=200$ foragers, $M=50$ targets in $1d$, $M=100$ targets in $2d$ and $\gamma_{max}=0.9$: (a) $P_{\rm T_{best}}$ as a function of the rate of memory use $q$ in one and two dimensions. (b) $P_{\rm T_{best}}$ as a function of the interaction parameter $\rho$ in two-dimensions for a fixed $q$. (c) Average number of neighbors at a distance $r$ of an individual in $2d$ ($r=7$). Inset: $1d$ case with $r=5$. (d) $P_{\rm T_{best}}$ as a function of the mean degree $\langle k\rangle$ in Erd$\ddot{\mbox{o}}$s-R\'enyi interaction networks. Inset: $1d$ case. (e) $P_{\rm T_{best}}$ in $2d$ as a function of the degree distribution exponent $\alpha$ in random scale-free interaction networks. The dotted line represent the complete graph value. Inset: $1d$ case. (f) $N_{inf}(k)$ is the average number of agents that reached for the first time one of the best resource sites (in $2d$) by using the experience of a node of degree $k$. The network is scale-free with exponent $\alpha=2$ and the degree distribution $N_k$ is shown with green triangles. We notice the overwhelming importance of highly connected nodes for transferring information. In all cases, averages over ten different landscapes are shown. 
  }}
\label{PtbestMult}
\end{figure*}

To investigate how the interacting walkers choose among many targets of different weights, we compare the computed (or perceived) ranking vs. the real ranking of the targets. The real ranking corresponds to the targets sorted according to their weights $\gamma_i$ (the best target has rank $1$) and the computed ranking refers to the targets sorted according to their final occupation probability $P_{T_i}$. If the group is able to { specifically select} the best targets, we expect a high correlation between both quantities for the best targets, and lesser correlations as the rank increases. As shown by Figs. \ref{rankP}(c)-(d), our walkers are not only capable of finding and exploiting these best targets, but overall, they compute the target rank very accurately. In our example, the $10\%$ best targets (the first 5 in $1d$ and the first 10 in $2d$) lie on the diagonal, meaning that these are always correctly identified as such by the group. However, fluctuations are strong for higher ranks. In particular, in $1d$, poorer targets are sometimes considered as important (occupied frequently), as many points lie under the diagonal.\\


{ Figure \ref{PtbestMult}(a) displays the variations with $q$ of the final occupation probability $P_{\rm T_{best}}$ defined as the probability that a walker chosen at random occupies the {\it best} target site (of highest $\gamma_i$). This quantifies here the foraging success of the group. $P_{\rm T_{best}}$ presents a maximum for a particular value of $q$. In $1d$, a value of $q=0.1$ is enough for the emerge of collective learning, or $P_{\rm T_{best}}$ large. We chose to keep $q=0.1$ in the $2d$ case for consistency and easier comparison. Slightly larger values ($q\sim 0.2$) lead to even stronger localization in $2d$. Figure \ref{PtbestMult}(b) displays $P_{\rm T_{best}}$ as a function of the interaction parameter $\rho$. A monotonous increasing behaviour is observed ({when $q>0$}) and a plateau} is reached at about $\rho=0.5$. In the following, we choose $\rho=0.5$. {When memory is suppressed, $P_{\rm T_{best}}$ is nearly vanishing for all $\rho$}.\\

These results suggest that many individuals aggregate around the best resource sites. We thus focus on the temporal evolution of the spatial cohesion of the group. 
To do so, we define the neighborhood of a forager by a segment of length $2r$ with $r=L/(2N)$ in $1d$, and by a disk of radius $r=L/(2\sqrt{N})$ in $2d$, centered at the forager position. The length $r$ represents half of the mean distance between two neighboring foragers if these would be distributed at random in space.  
For each forager, at time $t$, we determine the number of other group members located in its neighborhood, and take the average, denoted as $N_r$, over all foragers and independent simulations. An average over ten different landscape configurations is also performed.\\

{Figure \ref{PtbestMult}(c)} shows $N_r$ as function of time in $2d$. It can be noticed that no aggregation takes place when the walkers are memory-less ($q=0$) or when they have memory but do not interact ($q>0$ and $\rho=0$): in these cases each walker has, on average, only one neighbor ($N_r\approx 1$) at all times. In contrast, when the agents have both memory and the ability to transfer information ($q>0$ and $\rho>0$), $N_r$ increases with $t$ and reaches high values, specially in $1d$ {(Fig. \ref{PtbestMult}(c)- Inset)}. This result, together with the previous findings, indicates that the walkers aggregate around the best targets. Such selective aggregation is an outcome of collective learning { and indicates a high foraging success at the group level}.\\


We next investigate the effect of network connectivity among foragers on selective localization. For Erd$\ddot{\mbox{o}}$s-R\'enyi interaction networks, {Fig. \ref{PtbestMult}(d) displays the final $P_{\rm T_{best}}$} as a function of the mean number of connections per forager (mean degree), $\langle k\rangle=s(N-1)$, where $s\in[0,1]$ is the connectivity parameter. The { final} occupation probability of the best target is very low at small $\langle k\rangle$ but rapidly grows and saturates, both in $1d$ and $2d$. As soon as the walkers have on average 6 or 7 connections, { $P_{\rm T_{best}}$} has already the same value as for the fully connected graph ($\langle k\rangle=199$ here). We notice that at $\langle k\rangle\simeq6$, the giant component of the network is already well formed, since the percolation threshold occurs at $\langle k\rangle_c=1$ in random graphs \cite{barabasi16,newman10}. Therefore, collective learning requires much less connections than in a complete graph, but a network above the percolation threshold is not sufficient either.

Results for random scale-free networks with degree distribution exponent $\alpha$ (see ref. \cite{newman01} for details on their construction) are shown in {Fig. \ref{PtbestMult}(e)}.
The { final} occupation probability of the best target site is maximal for $\alpha=1$ (the minimum value of $\alpha$ owing to the normalization condition of the degree distribution), where it equals that of the complete graph. $P_{\rm T_{best}}$ decreases with $\alpha$, {\it i.e.} as the graphs become less heterogeneous. Note however that the mean connectivity also decreases with $\alpha$ in this network model. For $\alpha>3$, the networks are little connected and exhibit weaker localization. {Figure \ref{PtbestMult}(e)} shows that the most connected nodes (the "hubs") are critical for transferring information within the foraging group: the number of walkers that used the knowledge of a node of degree $k$ to arrive at one of the best target sites for the first time, $N_{inf}(k)$, is a function that increases sharply with $k$. Hence, highly connected nodes influence successful decisions of many other nodes.\\




\section{Discussion}
Memoryless random walk processes \cite{bart09, morales04, vis08, bart05, ben05, ben11} in some cases constitute advantageous strategies for searching for resources in unpredictable environments \cite{vis99}. Stochastic mobility models that incorporate memory are rarely solvable mathematically and their properties are less understood. An exception is the random walk with preferential relocations to places visited in the past, first introduced in \cite{gautestad05,gautestad06} and later solved in \cite{boyersolis,boyerpineda}. In this model, some displacements are completely random and local in space whereas others are based on memory and preferentially directed to sites that have been occupied often, sometimes far away from the forager's current position. 
Recently, we have shown that the trajectories generated by this model, so far studied in homogeneous environments, were actually able to localize specifically around some landscape heterogeneities representing resources \cite{falcon17}. { These sites are occupied for a longer time in each visit, compared to other places, and become rapidly reinforced after they have been visited for the first time. Therefore, a form of spatial learning emerges, reminiscent of a foraging animal adapting to its environment. Here we have studied systems of many agents connected by an information transfer network and following similar movement rules. One of our main results is that a large fraction of group members is able to sharply aggregate around the best resource site in a disordered environment with many resources.} { In the context considered here, this phenomenon represents a manifestation of collective learning.}\\

A distinctive property of collective learning revealed by the simulations is that individuals that exchange information { have a very high probability to occupying} the best resource site, among a large set of available resources. In contrast, when the walkers do not communicate, such selective localization does not take place and each walker learns only about its local environment, in the vicinity of the starting position. Therefore, both individual memory and information transfer between group members are necessary for successful group foraging via aggregation around the richest sites.\\

Aggregation phenomena { in populations of diffusive elements similar to the ones studied here} have been extensively studied in the context of taxis, where living organisms locally respond to external stimuli by modifying their movement statistics \cite{othmer97}. Large scale aggregation can be beneficial for the survival of starving bacterial colonies, for instance, as it facilitates the location and exploitation of food sources when these are scarce. Aggregation can occur by following diffusive chemicals or slime that individuals leave on their way. Theoretical models have been analysed mostly in completely homogeneous environments, without sites representing resources \cite{othmer97}. Stable aggregation patterns between the diffusive elements can occur in such cases after a spontaneous symmetry breaking, namely, the densities of population and external signal evolve in time toward non-uniform stationary patterns whose exact locations cannot be predicted exactly, since they depend in a complicated way on the initial density profiles. \\
%
%

In contrast, here we present one of the first studies (to our knowledge) on reinforced random walks in heterogeneous spaces, where translational invariance is broken. Whereas in the processes mentioned above, aggregation and self-localization do not occur at a specific location, in our model the walkers can aggregate in a given environment with a very high probability around the same few specific sites, independently of their starting positions. This property is reminiscent of the behaviour of ant colonies, where a vast majority of individuals is able to select, repeatedly over many experiments, the shorter of two paths separating the nest and a food resource \cite{bonabeau99}. { However, the mechanisms by which collective behaviours emerge in ants are related to chemical trail laying, and they differ from the information transfer processes of the current model, where relocations far beyond the current forager position are allowed.}
Our study shows that reinforced random walks can be useful for survival in complex environments, where a large number of foraging options exist. The problem studied here can be seen as an extension to spatial contexts of the basic stochastic models of reinforcement learning, where an agent chooses between a set of acts \cite{luce59,pemantle04}.\\

{ While our model has made the simplifying assumption that resources are always available, in real situations resource patches will be ephemeral, with a highly variable distribution and abundance \cite{sole99}. This implies that the best available patches, at any given time, will be changing constantly. 
Our results show that a group of foragers can locate the best available sites by performing reinforced walks and sharing information, but they do not actually find all available resources, or forage in all of them according to their rank. This may also imply that in realistic situations, resource use may not always be optimally efficient \cite{pyke84}, especially at the individual level \cite{seth01}.} { Competition for resources should also be considered in future modelling, since it represents a cost of information transfer. For example, competition could decrease cohesion and cause the formation of sub-units exploiting a wider range of resource sites \cite{miramontes16}.}
\\

Our model highlights the importance of the topology of the interaction network on the group's ability for learning collectively. When the interaction network is of Erd$\ddot{\mbox{o}}$s-Renyi type, a modest average number of neighbours per individual (six or seven) is sufficient to trigger collective aggregation around the best resource sites. Further increase in the connectivity does not lead to stronger aggregation/cohesion. This characteristic number of connections is nevertheless significantly larger than the percolation threshold, given by $\langle k\rangle =1$ neighbour for ER networks \cite{newman10}. It means that the networks that are efficient for collective learning not only have a giant component but are also densely connected (although not complete), as observed in many primate groups \cite{kasper09}. In spider monkey social networks, in particular, the sum of the association indices of a node (equivalent to its strength) varies between 2 and 7, depending on the individual and year of study \cite{ramosfer09}. Values around 5 have been reported in other animals, such as dolphins \cite{lusseau04} or giraffes \cite{carter13}.\\

We also found that the presence of highly connected individuals (or "hubs") in a random, scale-free network is important for collective learning. This is coincident with the results of an empirical study of spider monkeys \cite{palacios_inrev} that found that more central individuals lead collective movements and are followed to available feeding trees that they know about more frequently than non-central individuals.\\

In flocks of birds, each individual interacts on average with about 6 nearest neighbours independently of the separation distance to these individuals \cite{ballerini08,young13}. Such topological interactions contribute to the maintenance of robust collective motion and cohesiveness, even in the presence of external perturbations. In flocks, a fixed number of interactions is thought to optimize the trade-off between group cohesion and individual effort \cite{young13}. Similarly, in foraging groups, six or seven neighbours may be a good compromise between the benefit of gathering valuable information from others and limited sensory and memory capacities. In an empirical study aimed specifically at uncovering the rules by which spider monkeys share information about feeding trees \cite{palacios_inrev}, individuals that know about available feeding trees share this information with an average of 4.25 ($\pm 3.75$ S.D.) individuals over several detailed observations at feeding trees. More studies are needed on the generality of these minimum numbers of connections, in particular with respect to variations in foraging group size.\\

{ Our results allow us to make some predictions for future studies of collective learning in animals. First, groups of animals with the ability to remember the location of resource sites will forage more efficiently when they can exchange information about the location of these sites. This could be tested by manipulating or recording the location of new sites and monitoring the transfer of information between individuals, as in \cite{palacios_inrev}. Second, the number of interaction partners, as well as the configuration of the information transfer network will be important for collective learning to emerge. This could be tested experimentally by manipulating the number of partners each individual has access to, and testing whether the minimum number of connections mentioned above is relevant for real networks.}\\

The memory-based algorithm developed in the present work might find applications in swarm robotics. Experiments in this area can help to better understand collective decision making in nature \cite{bose17}. In general, a large number of autonomous robots can be designed to coordinate with each other to perform a common task via exchange of information { with other robots within} a local range. The dynamical rules based on memory studied here require little computation and our results suggest that opinion aggregation could be enhanced thanks to learning processes. The results for 2D disordered environments also suggest that swarms with memory represent a promising tool to solve complex optimization problems in { high} dimensional spaces.\\

To summarize, { we have presented a model that exhibits} collective learning due { to the combined} use of memory and information transfer { between agents}. { The foraging group as a whole is} capable of distinguishing the best resource sites and localize around them. While the specific topology of the network through which they share information seems to not have great impact in these emerging features, it is enough that each walker has, on average, six or seven neighbors (or that there are enough highly connected individuals) for  collective learning to appear. This model provides hypotheses and predictions for empirical studies of collective learning, as well as suggesting design principles for search algorithms and pattern recognition.\\
 
 Acknowledgements:  DB acknowledges support from DGAPA-PAPIIT UNAM Grant No. IN105015. AFC thanks { A. Aldana for fruitful discussions and} Conacyt (Mexico) and PAEP-UNAM for financial support. {GRF acknowledges C3-UNAM for logistical support and National Geographic (grant WW008R17) for travel expenses.}

\end{document}